\documentclass[aps,preprint,showpacs]{revtex4}%
\usepackage{amsfonts}
\usepackage{amsmath}
\usepackage{amssymb}
\usepackage[dvips]{graphicx}
\providecommand{\U}[1]{\protect\rule{.1in}{.1in}}

\begin{document}
\preprint{ }
\title[Duality violation test]{Experimental test for violation of duality on a photon beam}
\author{Daniel Mirell}
\affiliation{Department of Physics and Astronomy, University of New Mexico, Albuquerque, NM 87131}
\author{Stuart Mirell}
\affiliation{Cyclotron Facility, Bldg. 345, VA GLAHS, Los Angeles, CA 90073 and Department
of Radiological Sciences, University of California at Los Angeles, Los
Angeles, CA 90095}

\begin{abstract}
We test for evidence violating the duality invariant ratio of
photon beam irradiance and wave intensity. Split beams from a 633nm HeNe laser
are intersected at a diffraction grating complementary to the resultant
interference pattern. An output beam from the grating, depleted in irradiance
relative to wave intensity from the perspective of local realism, is
transiently intersected with a beam from an independent HeNe laser and
measured irradiance is amplified by $\sim$4\% in conflict with quantum mechanics.

\end{abstract}
\volumeyear{ }
\volumenumber{ }
\issuenumber{ }
\eid{ }
\date{September 3, 2005}
\startpage{1}
\endpage{17}
\maketitle

\section{Introduction}

A basic representation of local realism presented earlier postulates that
photons are comprised of separate real entities of wave structure and energy
quanta [1, 2]. In this representation, the wave intensity is a relative
probability density that determines the distribution of the resident energy
quanta as a proportionate energy flux density consistent with Born's first
principle. A wave entity is independent of its resident energy quanta, and
wave entities are mutually non-interactive. These properties are essentially
equivalent to those of the locally real wave structures postulated by de
Broglie \cite{debroglie} and interpreted as propensities by
Popper\ \cite{popper}. Interference is a consequence of the superposition of
these non-interactive wave entities in a region of mutual intersection. The
physical manifestation of this interference in the intersection region is the
redistribution of the energy flux density proportionate to the superposition
wave intensity.

Conversely, in quantum mechanics photons are treated as probabilistic entities
that exhibit dualistic wave or particle properties depending upon the nature
of the measurement process. The proportionality between wave intensity and
energy flux density is a superfluous invariant. Moreover, a photon can
interfere only with itself.

Then the class of experiments involving two intersecting independent photon
beams seemingly provides a testable basis for differentiating between local
realism and quantum mechanics. Many such experiments have demonstrated that
interference does occur in apparent conflict with quantum mechanics
\cite{hull}.

However, Mandel argued that the results of these experiments do not constitute
a violation of quantum mechanics since, for any photon, we do not know on
which beam that photon had initially resided \cite{mandel}. This lack of
knowledge provides a loophole that allows any photon, with its initial beam of
residence not known, to interfere with itself. Consequently, because of that
loophole, this class of experiments is not generally viewed as differentially
testing quantum mechanics and local realism. Mandel further contended that
this conclusion was applicable to the discrete realm as well as to the
continuous wave realm.

Our objective here is to close Mandel's loophole by taking this class of
experiments employing intersecting independent beams an additional step. In
the continuous wave realm, we prepare one of the beams in accord with the
locally real representation such that the beam is in a state depleted in
energy flux density (irradiance) relative to the beam's wave intensity noting
that this property of depletion is contrary to quantum mechanics
\cite{mirell3}. A transient spatial coupling of an independent beam and that
depleted beam provides for a zone of mutual interference of those beams. We
still do not know for any given photon on which beam that particular photon
had initially resided so we certainly continue to expect interference to occur
in this zone from the perspectives of both local realism and quantum
mechanics. However, for local realism the exchange of energy quanta is not
completely random. A potentially measurable net redistribution of energy
quanta should occur onto the depleted beam from the independent beam in an
equilibration process. This net redistribution of course is not consistent
with quantum mechanics since depletion itself is excluded.

\section{Principles}

In the present work, reference to \textquotedblleft quantum
mechanics\textquotedblright\ implicitly denotes the standard probabilistic
interpretation of that physical representation. Conversely, \textquotedblleft
local realism\textquotedblright\ implies a physical representation devoid of
the particular conflicting tenets of the probabilistic interpretation but
still fundamentally consistent with the quantum mechanical formalism.

Because these two representations have distinct similarities and differences,
it is essential that we precisely define several critical quantities. For some
point on the cross section of a given beam the energy flux density, identified
as the irradiance $I$, is measurable by a conventional energy sensitive
detector. The associated wave intensity $W$, which is effectively a
probability density, may be measured by assessing interference visibility with
a reference beam. $I$ and $W$ vary proportionately across the beam's cross
section consistent with Born's first principle. The ``occupation value'' of a
beam defined as
\begin{equation}
\Omega=\frac{I}{W}%
\end{equation}
is then constant at all points on a beam's cross section. Integrating $I$ over
a beam cross section $\int I\,da$ gives the total beam power while the
analogous integral $\int W\,da$ is recognized as the total probability. Since
$I$ and $W$ vary proportionately, the ratio of their integrals also yields the
same value of $\Omega$. Nevertheless, it will generally be more convenient to
work with the quantities $I$ and $W$ rather than their respective integrals.
Moreover, $I$ and $W$ can usually be treated as representing the respective
maximum values on a particular beam cross section e.g. for a Gaussian profile,
the values at the geometrical center.

In the present context, the two representations fundamentally differ in the
treatment of wave intensity $W$. For quantum mechanics $W$ is an absolute
probability density that is in a fixed proportion to $I$. Conversely, for
local realism $W$ is a relative probability density. The immediate consequence
of this disparity is that for quantum mechanics the occupation value is
inherently some constant $\Omega_{o}$, rendering that quantity as superfluous,
whereas for local realism a ``non-ordinary'' beam could, in principle, be
prepared such that $\Omega\neq\Omega_{o}$ (although we expect that virtually
all typically encountered beams will be ``ordinary'' with an occupation value
$\Omega_{o}$ not in conflict with quantum mechanics). Arbitrary units can
always be employed such that $I$ and $W$ are pure numerical values mutually
scaled to equality. Consequently, for this choice of units the occupation
value $\Omega_{o}=1$ is universal for quantum mechanics but is applicable to
local realism only for ordinary beams. Then, relative to resident energy
quanta, a beam for which $\Omega<1$ is appropriately defined as ``depleted''.

In the discrete realm, where at most only an individual photon is present on
any macroscopic segment of beam path, an $\Omega$ for local realism may
deviate from unity by means as trivial as a simple beam splitter. For the
discrete realm, as well as the continuous wave realm, the incident wave
intensity is always fractionally divided between the two output channels in
accord with the transmission and reflection coefficients of the beam splitter,
but, most noticeably for discrete events, the particular output channel
assumed by the single energy quantum is random and uncontrollable, albeit,
statistically predictable at the exit face of the beam splitter. Then, for a
particular output channel and event, the resultant output beam is also
randomly and uncontrollably transiently ``depleted'' or ``enriched'' in energy
quantum relative to the wave intensity.

Quantum mechanics is not violated in this discrete phenomenon since
probabilistically, on both of the output beams, we still have an expectation
of an energy quantum proportionate to the fractional wave intensity.

Historically, many investigations designed to test the reality of empty
(totally depleted) de Broglie waves have used related methods. However these
methods can only yield the very weak wave intensities associated with discrete
photon beams.

We report here the results of an experimental test of this disparity between
quantum mechanics and local realism using an apparatus that should generate
depleted beams in the continuous wave realm from the perspective of local
realism. A transient equilibration of one of these beams with an ordinary beam
is then expected to provide evidence resolving that disparity.

As a prelude to considering the details of the apparatus, we first consider a
single beam incident on two similar but critically distinctive gratings
$G_{u}$ and $G_{c}$ (Max Levy Autograph, Inc., BA011 and ZY002, respectively).
Both have the basic structure of a Ronchi ruling, an array of equal width
transmissive and opaque bands, with opaque bands at 100 per inch on a glass
plate. The opaque bands on $G_{u}$ are thin $\sim0.1\mu m$ depositions of
black evaporated chrome whereas on $G_{c}$ the opaque bands are fabricated by
etching the glass support plate to a depth of $\sim10\mu m $ and filling with
black epoxy. The resultant diffractive orders of the two gratings differ
measureably in relative irradiances. Most notably the irradiance ratio of the
$0^{th}$ order relative to the $\pm1^{st}$ order is $1:0.41$ for $G_{u}$ and
$1:0.57$ for $G_{c}$. This deviation is not unexpected since the relative
irradiances of diffractive orders are a very sensitive function of the
periodic structures.

For an idealized Ronchi ruling, the relative irradiance of the $i^{th}$ order
is given by $sinc^{2}(i\pi/2)$. The $0^{th}$ to $1^{st}$ order ratio,
$1:0.405$, is very nearly equivalent to that of $G_{u}$. Accordingly, we
approximate $G_{u}$ as an idealized Ronchi ruling with relative wave
amplitudes given by $sinc(i\pi/2)$ which is used to generate a normalized set
of amplitudes $\{A_{ui}\}=$ 0.506, 0.323, 0, -0.107, 0, 0.064, 0, -0.046 for
the $\pm i^{th}$\ order $0^{th},\pm1^{st},\pm2^{nd}$ etc. truncated at
$\pm7^{th}$ since higher orders are substantially diminished. In the scalar
treatment applicable to this relatively coarse grating, the squares of these
amplitudes $A_{ui}^{2}=W_{ui}=I_{ui}$ where $W_{ui}$ and $I_{ui}$\ are
respectively the $\pm i^{th}$ order wave intensity and irradiance for a single
beam incident on $G_{u}$. The set $\{A_{ui}\}$ is normalized here such that
the sum of squares gives a 0.5 transmitted wave intensity and\ a 0.5
transmitted irradiance where the incident $W=I=1$. This normalization is
consistent with the 50\% transmissivity of Ronchi rulings to plane radiation.
Because of the equivalence of transmitted irradiance and wave intensity for a
single incident beam, the theoretical occupation value computed for the total
transmitted set is trivially $\Omega_{uT-th}(1)=0.5/0.5=1$ where the argument
specifies that one beam is incident. Equilibration further provides that the
occupation value $\Omega_{ui-th}(1)=1$ for each individual transmitted
$i^{th}$ order diffraction beam since the energy quanta at the exit face of
the grating distribute onto the individual beams in proportion to their
respective wave intensities. Accordingly, we suppress the subscripts $T$ and
$i$, and $\Omega_{u-th}(1)=1 $ is understood to apply to both the total and
the individual beams of the diffraction orders.

To first order, the higher irradiance ratio for $G_{c}$ relative to that of
$G_{u}$ is equivalent to a relative broadening of the diffractive envelope.
(This broadening could equivalently be generated by providing $G_{u}$ with
with plano-concave transmissive bands.) The higher orders of $G_{c}$ can be
approximated by numerically fitting the $sinc$ argument of $G_{u}$ to yield
the ratio $1:0.57$ which results in $i\pi/2\rightarrow0.8i\pi/2$. The
corresponding set of $G_{c}$ normalized wave amplitudes is $\{A_{ci}\}=$
0.459, 0.347, 0.107, -0.072, -0.086, 0, 0.058, 0.031 for orders $0^{th}%
,\pm1^{st},\pm2^{nd}$ etc. again truncated at $\pm7^{th}$. This set, like that
of $G_{u}$, is also normalized such that the sum of squares gives a 0.5
transmitted wave intensity and, equivalently, a 0.5 transmitted irradiance for
the chosen normalization of the incident beam. Similarly, for a single
incident beam, the theoretical occupation value computed for the total
transmitted set is trivially $\Omega_{c-th}(1)=1$ and this value also applies
to each individual transmitted beam. Then both $\Omega_{u-th}(1)$ and
$\Omega_{c-th}(1)$ are unremarkably unit valued consistent with quantum
mechanics and local realism.

With the above set of amplitudes for $G_{u}$ and for $G_{c}$, we can proceed
to calculations of the respective $\Omega_{u-th}(2)$ and$\ \Omega_{c-th}(2) $,
theoretically predicted by local realism for two appropriately aligned
incident beams. These calculations are intended here purely as an exercise in
providing locally real predictions for the experimentally measured
counterparts $\Omega_{u-exp}(2)$ and $\Omega_{c-exp}(2)$.

The two mutually coherent, mutually converging incident beams are angularly
separated at some $\theta_{i}$ such that the resultant interference pattern
has a spatial periodicity equal to that of the grating. For a relatively
coarse grating, such as a Ronchi ruling with 100 opaque bands per inch, the
small angle approximation and a scalar treatment of amplitudes are applicable.
For the above choice of $\theta_{i}$,\ the angular separation of the adjacent
diffraction orders from either incident beam $\theta_{d}=\theta_{i}$ resulting
in spatial coincidences of orders from the two incident beams.

We can treat the gratings as two-stage components, a periodic array of opaque
and transmissive bands followed by some diffractive structure associated with
the transmissive bands. The calculation of $\Omega_{u-th}(2) $ and$\ \Omega
_{c-th}(2)$ requires a determination of the total transmitted irradiance and
the total transmitted wave intensity for each.

The total transmitted irradiance
\begin{align}
I_{T}(2)  &  =2\frac{\int_{-\frac{\pi}{4}}^{\frac{\pi}{4}}\cos^{2}x\;dx}%
{\int_{-\frac{\pi}{2}}^{\frac{\pi}{2}}\cos^{2}x\;dx}\nonumber\\
&  =1.64
\end{align}
is simply the integrated peak-centered 50\% of each interference pattern cycle
normalized to the entire cycle ($\Delta x=\pi$). The leading factor of two is
the total incident irradiance of which 82.5\% is transmitted where $I$ and $W$
for both incident beams are normalized to unity. $I_{T}(2)$ is the same for
both gratings since they share the same first stage array of opaque and
transmissive bands.

The second stage total wave intensity for $G_{u}$ is given by
\begin{align}
W_{uT}(2)  &  =\sum_{i=-n}^{n+1}(A_{ui}+A_{ui-1})^{2}\nonumber\\
&  =1.66
\end{align}
where $A_{ui}$ and $A_{ui-1}$ in each term are understood to be the coincident
amplitudes produced by the respective incident beams. The amplitudes for
$G_{u}$ were given above to $\pm7^{th}$ order and since their magnitudes
diminish substantially by that order the series is truncated by imposing
$A_{ui}\equiv0$ for $\left|  i\right|  >7$ and $n=7$.

The resultant occupation value
\begin{align}
\Omega_{u-th}(2)  &  =\frac{I_{T}(2)}{W_{uT}(2)}\nonumber\\
&  =0.99\nonumber\\
&  \approx1
\end{align}
strongly suggests that the configuration of two beams incident on the
idealized grating $G_{u}$ does not generate output beams that are in conflict
with quantum mechanics.

As a final exercise, we similarly calculate $\Omega_{c-th}(2)$. As before,
$I_{T}(2)=1.65$. The total wave intensity
\begin{align}
W_{cT}(2)  &  =\sum_{i=-n}^{n+1}(A_{ci}+A_{ci-1})^{2}\nonumber\\
&  =1.79
\end{align}
is recognized as a summation functionally identical to Eq. (3) but using the
$G_{c}$ single beam amplitude set $\{A_{ci}\}$ given above with $A_{ci}%
\equiv0$ for $\left|  i\right|  >7$ and $n=7$. The resultant $W_{cT}(2)$
yields an
\begin{align}
\Omega_{c-th}(2)  &  =\frac{I_{T}(2)}{W_{cT}(2)}\nonumber\\
&  =0.92.
\end{align}

The significant deviation of $\Omega_{c-th}(2)$ from unity predicts that the
two beam configuration on $G_{c}$, unlike that on $G_{u}$, presents a conflict
between quantum mechanics and local realism. Quantum mechanics, of course,
would require that $\Omega_{c-th}(2)\equiv1$ facilitated in the present case
by a renormalization of the wave (probability) amplitudes. This
renormalization necessarily must occur in $G_{c}$ at the ``first stage''
temporally coincident with the partial collapse of amplitudes at the opaque
bands since no further collapse occurs up to and at the totally transmissive
``second stage''. The renormalization must decrease the total transmitted
probability density in order to maintain the quantum mechanically required
unit value of $\Omega_{c-th}(2)$.

Quantum mechanics then, quite curiously, arbitrarily requires a probability
amplitude renormalization when the grating is $G_{c}$ but not when the grating
is $G_{u}$. Moreover, when a strict interpretation of quantum mechanics is
imposed, an even more bizarre phenomenon is evident. With the gratings treated
as two-stage components, amplitude collapse and any renormalization must occur
at a time $t=0$ when the incident wave encounters the first stage. At some
later time $t=\tau$, the wave encounters the second stage of either $G_{u}$ or
$G_{c}$. Since only the second stages of these gratings are distinguishable,
the amplitude collapse and any renormalization of the wave must have
anticipated at $t=0$ which of those two gratings is in place. The implications
of this anticipation go beyond even that of the non-local response of
separated entangled photons or particles. However, it is not our intent here
to examine the quantum mechanical consequences of the foregoing exercise in
theoretical grating experiments, rather this exercise provides a motivation
for constructing an apparatus with $G_{c}$ and experimentally testing quantum
mechanics against local realism based upon the disparate predictions of
$\Omega_{c-th}(2)$.

\section{Apparatus and Procedures}

The experimental apparatus for this test is shown in Fig. 1. Aside from the
lasers, all essential components and optical paths are depicted omitting only
mirrors used for aligning and folding those paths. A 4mW polarized 633 nm HeNe
laser generates an ordinary beam $S$ (for which $\Omega_{o}=1$) that is
transmitted by a 0.7 transmissive beam splitter $BS\,1$ to lens $L\,1 $
($f=100$ $mm$) and into a diffractive beam splitter $DBS$ (Mems Optical 1019,
fused silica 16 beam splitter). An aperture $Ap\,1$ restricts the output of
that beam splitter to two equal irradiance beams $S_{1}$ and $S_{2} $ mutually
diverging at 6.5 mrad toward lens $L\,2$ ($f=150$ $mm$). $L\,2$ converges the
two beams at the plane of grating $G_{c}$ with a periodic structure width
$b_{g}=0.254$ $mm$. The $S_{1}$ and $S_{2}$ individual beam self-divergences
are slightly reduced relative to that of the initial $S$ beam by the combined
action of $L\,1$ and $L\,2$ serving as a beam expander. More significantly,
this beam expander, straddling the diffractive beam splitter, avoids large
self-divergences of $S_{1}$ and $S_{2}$ that would otherwise result if $L\,1$
were omitted.

\begin{figure}[htb]
\special{isoscale Mirellduality587.eps, \the\hsize 3in}
\vspace{3in}
\includegraphics[height=1.3154in,width=6.1272in]{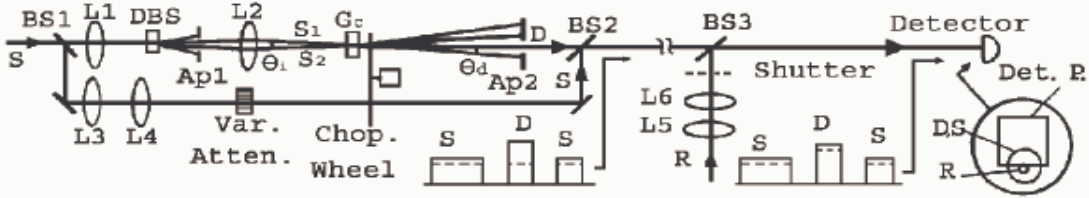}
\caption{Experimental setup, not to scale, showing transmitted
depleted $D$ pulse, received equilibrated $D$ pulse, and frontal
detail of beam spots incident on detector photodiode.}
\end{figure}%

The $S_{1},S_{2}$ beamspot at grating $G_{c}$ is characterized by a linear
array pattern of interference fringes with maxima separated by $b_{i}%
=\lambda/\theta_{i}$ in the small angle approximation. Shifting the positions
of the rail-mounted diffractive beam splitter $DBS$ and grating $G_{c}$ along
the optical axis permits adjustment of $\theta_{i}$ such that the resultant
final interference pattern matches the grating's periodicity, i.e.
$b_{i}=b_{g}=0.254\,mm$, while maintaining beam spot coincidence of $S_{1}$
and $S_{2}$ at $G_{c}$. Concurrently, $\theta_{i}=\theta_{d}$ with this
adjustment. A micro-translation stage is used to laterally shift $G_{c}$ with
respect to the interference pattern thereby centering the principal maxima
over the transmissive bands of the grating and maximizing the transmitted
irradiance. The output beams of $G_{c}$ are intermittently blocked or
transmitted by an adjacent chopper wheel.

From the perspective of quantum mechanics and local realism, we expect that
the total irradiance emerging from $G_{c}$ will distribute onto the output
beams in proportion to the respective wave intensities given as terms in the
Eq. (3) summation. For local realism, these output beam wave intensities
represent real entities on the field that provide ``relative'' probabilities
in the distribution of total transmitted irradiance onto the output beams. For
clarity, only four of these beams are depicted in the figure. The two
center-most beams, both with $(A_{c1}+A_{c0})^{2}$ wave intensities, will have
the highest irradiance of the output beams. An aperture $Ap\,2$ transmits one
of these two, designated as beam $D$, and blocks all other $G_{c}$ output
beams. This designation is given since local realism predicts that $D$ is
depleted. $D$, which is a pulsed beam as a consequence of the chopper wheel,
propagates to a 0.7 transmissive beam splitter $BS\,2$.

Concurrently, when $S$ encounters $BS\,1$, a fractional portion is reflected.
That portion, still designated as $S$, is directed through a beam expander
comprised of lenses $L\,3$ ($f=100$ $mm$) and $L\,4$ ($f=150$ $mm$), into a
variable attenuator (Edmund Optics G41-960 mounted on a lateral translation
stage) and through the chopper wheel to produce a pulsed $S$ beam that also
propagates to $BS\,2$.

Two concentric sets of wheel apertures are configured such that $D$ and $S$
pulses emerge non-simultaneously. The two pulsed beams are incident at a
common point on beam splitter $BS\,2$ where they are angularly combined and
``transmitted'' as a single interlaced $D,S$ pulsed beam .

That pulsed beam is ``received'' at a 1000 mm distant 0.7 transmissive beam
splitter $BS\,3$. An independent 3 mW HeNe laser generates a constant
irradiance beam $R$ that enters a beam expander, $L\,5\ $($f=100$ $mm$) and
$L\,6$ ($f=200$ $mm$). Passage of $R$ is controlled by a shutter before
reaching a common incidence point with the pulsed $D,S$ beam on $BS\,3$. The
trajectory of the reflected $R$ is adjusted by $BS\,3$ to very nearly coincide
within $\lesssim1$ mrad with that of the transmitted pulsed $D,S$ beam such
that at the terminus of a 3000 mm ``coupling'' path, the $R$ beam spot is
displaced to one side of the common $D,S$ beam spot as shown in the detail on
Fig. 1. The $L\,5,L\,6$ beam expander is deliberately set to give $R$ a slight
convergence that positions the minimum waist at the coupling path terminus.
These settings are intended to maximize irradiance equilibration of the $D$
and $S$ pulses with $R$ over the 3000 mm coupling path while reducing the $R$
beam spot on the common $D,S$ beam spot at the path terminus. This facilitates
selective positioning of the beam spots across the edge of the (Edmund Optics
G54-038) detector's 100 mm$^{2}$ Si photodiode such that a large fraction of
the common $D,S$ beam spot is incident on the photodiode but the $R$ beam spot
is largely excluded.

At the terminus, with the beam spots temporarily shifted to the center of the
photodiode, the observed power of the total non-pulsed $D$ beam spot is $\int
I_{Di}\,da\approx20$ $\mu W$. Here the subscript ``$i$'' denotes $I_{Di}$ as
the initial $D$ irradiance, i.e. before coupling with $R$. The variable
attenuator allows the $S$ beam spot to be adjusted to a comparable power.
Additionally, the $L3,L4$ beam expander is set so that the $S$ beam spot has
the same irradiance profile as that of the $D$ beam spot. The total observed
$R$ beam spot power is $\int I_{R}\,da\approx180$ $\mu W$. However, when
acquiring the experimental data, selective positioning of the beam spots on
the edge of the photodiode, as shown in the Fig. 1 detail, results in
interception of fractions $F_{D}\approx0.4$ of the pulsed $D$ (and $S$) beam
spot power and $F_{R}\approx0.02$ of the constant $R$ beam spot power. The
amplified detector output is routed to an oscilloscope (Tektronix TDS 1002).
The\ $S_{1},S_{2}$ interference maxima are centered on the $G_{c}$
transmissive bands by adjusting the micro-translation stage to give maximum
detected irradiance of the $D$ beam with the chopper wheel stationary and the
$R$ beam blocked by the shutter.

The chopper wheel is then set into\ rotation repetitively transmitting a
sequence of an 8 msec $S$ pulse, a 4 msec blank, a 4 msec $D$ pulse, a 4 msec
blank, a 4 msec $S$ pulse, and an 8 msec blank. The longer, first $S$ pulse
provides for oscilloscope triggering based on pulse width. The oscilloscope is
operated with a 128 trigger cycle averaging to increase measurement precision.
Pulse width triggering is set to $\geq$ 6 msec to ensure synchronization with
the 8 msec. $S$ pulses. The measured pulse heights of the 4 msec $D$ and $S$
pulses are denoted as $P_{Di}$ and $P_{Si}$, respectively. The shutter is then
opened, unblocking $R$, and after a 30 sec. interval to allow for equilibrium
of the 128 averaging, pulse heights denoted as $P_{Df}$ and $P_{Sf}$ are
measured. Each complete trial consists of a set of these four pulse height
measurements, $P_{Di}$, $P_{Si}$, $P_{Df} $ and $P_{Sf}$. An experimental
``run'' consists of a series of these trials. As a final calibration procedure
before each such run, $R$ is blocked and the variable attenuator is adjusted
to give $P_{Si}=P_{Di}$.

\section{Results and Analysis}

The essential data of each experimental trial resolves to pulse height ratios
$P_{Di}/P_{Df}$ and $P_{Si}/P_{Sf}$. We can identify
\begin{equation}
\Omega_{Di}=P_{Di}/P_{Df}%
\end{equation}
as a single trial occupation value of the $D$ pulses prior to their
equilibration with the $R$ beam. This follows from $P_{Di}/P_{Df}%
=I_{Di}/I_{Df}$ where $I_{Df}=W_{D}$ assuming that the $D$ pulses are fully
equilibrated after coupling with the $R$ beam. For the $S$ pulses in each
trial, the ratio $P_{Si}/P_{Sf}\approx1$ serves as an ``experimental control''
for the accompanying $D$ pulse height ratio $P_{Di}/P_{Df}$. (The pulse
heights of the 8 msec $S$ trigger pulses were not distinguishable from those
of the accompanying 4 msec $S$ pulses used to provide trial values.) The
$P_{Si}/P_{Sf}$ ratios are typically tightly clustered about unity consistent
with expected detector measurement error and laser power fluctuations.
Initially setting $P_{Si}=P_{Di}$ ensures that the detector response is highly
linear in measuring $P_{Di}$ relative to $P_{Df}$ despite the presence of an
$R$-produced baseline in the latter measurement provided that the control
condition $P_{Si}/P_{Sf}\approx1$ is satisfied.

Results are given here for a representative experimental run consisting of a
series of 40 trials. With the oscilloscope operating on 128 trigger cycle
averaging, this run comprises a total of $\sim10^{4}$ $D$ pulses. The
trial-averaged ratio $\left\langle P_{Di}/P_{Df}\right\rangle =\left\langle
\Omega_{Di}\right\rangle =\Omega_{c-exp}(2)=0.96\pm0.01$ where $\Omega
_{c-exp}(2)$ is the experimental counterpart to $\Omega_{c-th}(2)=0.92$,
theoretically predicted by local realism. The accompanying trial-averaged
control ratio $\left\langle P_{Si}/P_{Sf}\right\rangle =1.00\pm0.01$ is
consistent with both quantum mechanics and local realism.

The results for the pre- and post-coupling configurations are schematically
shown in the Fig. 1 pulse sequences where the irradiance $I$ and the wave
intensity $W$ are respectively depicted as dashed and solid lines. Quantum
mechanically, the transmitted $D$ pulses should not acquire net irradiance
from coupling with the $R$ beam. However, for local realism an $\Omega
_{c-exp}(2)=0.96$, i.e. $\approx4\%$ depletion, is well within the range of
expected values.

It is of some particular corroborative interest that experimental data
acquired with the grating $G_{c}$ replaced by the ``idealized'' grating
$G_{u}$ yields an $\Omega_{u-exp}(2)$ not statistically different from unity.
This result is consistent with the Eq. (4) theoretical prediction of
$\Omega_{u-th}(2)\approx1$ and constitutes a significant experimental control.

Potential sources of errors have been examined to determine their impact on
the measurement of $\Omega_{c-exp}(2)$. With regard to local realism, the
assumption implicit in Eq. (7) that the final irradiance $I_{Df}$ of the $D$
pulses is completely equilibrated constitutes one probable source of error. In
general, local realism predicts an equilibration between two transiently
coupled beams of different $\Omega$ that results in a net transferrence of
irradiance. That irradiance transferrence is directly analogous to the charge
transferrence of coupled capacitances. An initial and final $\Omega$ and $\int
I\,da\,$(power) of these beams with total relative probabilities $\int
W_{D}\,da$ and $\int W_{R}\,da$ on beams $D$\ and $R$ respectively are
calculationally equivalent to the initial and final $V$ and $Q$ of two
capacitances $C_{D}$ and $C_{R}$ transiently coupled $+$ to $+$ and $-$ to $-
$. Following this analog, the final beam power on $D$ is
\begin{align}
\int I_{Df}\,da  &  =(\int I_{Di}\,da+\int I_{Ri}da)\frac{\int W_{D}\,da}{\int
W_{D}\,da+\int W_{R}\,da}\nonumber\\
&  \approx\int W_{D}\,da
\end{align}
if $\int W_{R}\,da\gg\int W_{D}\,da$, $\int I_{Ri}\,da\gg\int I_{Di}\,da$, and
where arbitrary units provide $\int I_{Ri}\,da/\int W_{R}\,da=\Omega
_{Ri}=\Omega_{o}=1$. Consequently,
\begin{align}
\Omega_{Df}  &  =\frac{\int I_{Df}\,da}{\int W_{D}\,da}\nonumber\\
&  \approx\Omega_{o}\nonumber\\
&  =1
\end{align}
with the $R$ beam serving as a nearly infinite irradiance ``source''
negligibly altered from its initial ordinary beam occupation value by the
transfer of irradiance in the coupling process. These conditions should be
reasonably approximated given the ratio of total beam powers $\int
I_{Ri}\,da/\int I_{Di}\,da\approx180\mu W/20\mu W$.

Completeness of equilibration is also dependent upon the coupling efficiency.
Statistically significant $\left\langle P_{Di}/P_{Df}\right\rangle <1$ are
measured when the mutual divergence angle of the pulsed $D,S$ and $R$ beams is
$\sim0.8$ mrad. As this angle is increased to $\gtrsim1.2$ mrad, the resultant
$\left\langle P_{Di}/P_{Df}\right\rangle $ plateaus at $\approx1$.
Accordingly, the mutual divergence angle on the $\sim3000$ mm overlapping
$D,S$ and $R$ beam paths must be carefully aligned and maintained for the
duration of a series of trials. Mechanical instability, transiently increasing
this angle, is expected to produce trials for which $P_{Di}/P_{Df}%
\rightarrow1$ while concurrently the control value $P_{Si}/P_{Sf}$, remaining
at $\approx1$, would not provide an indication of these transient angular
increases. A minimal dispersion of $\left\langle P_{Di}/P_{Df}\right\rangle $
is probably the most reliable indicator of stable coupling alignment for a
given series of trials.

In any case, regardless of the origin of incomplete equilibration, the
experimentally measured $\Omega_{c-exp}(2)$ would be closer to unity than the
actual occupation value and, equivalently, the actual depletion would be underestimated.

We next consider a source of error associated with the fractional presence of
the $R$ beam spot on the detector during coupling. Eq. (7), aside from
equilibration assumptions, further assumes that $P_{Di}$ and $P_{Df}$ are
proportional measures of the respective $D$ beam irradiances $I_{Di}$ and
$I_{Df}$ incident on the detector. (The proportionality would also extend to
the incident powers $\int F_{D}I_{Di}\,da$ and $\int F_{D}I_{Df}\,da$.) The
assumption of proportionality is certainly reasonable for $P_{Di}$ and would
seem to be equally reasonable for $P_{Df}$ since the latter quantity is
measured relative to the baseline generated by the constant $R$ beam. However,
on closer inspection the proportionality is not exact from the viewpoint of
local realism and we need to examine the magnitude and sign of the discrepancy.

During coupling, the $D$ beam power increases by
\begin{equation}
\Delta=\int(I_{Df}-I_{Di})da
\end{equation}
but this gain must equal the loss of power on the $R$ beam
\begin{equation}
\Delta=\int(I_{Ri}-I_{Rf})da.
\end{equation}
Then during coupling, i.e. while the $D$ pulse is actually present on the beam
path, the supposedly constant baseline of the $R$ power incident on the
detector actually drops by $F_{R}\Delta$ and the apparent pulse height
\begin{equation}
P_{Df}\varpropto\int F_{D}I_{Df}\,da-F_{R}\Delta
\end{equation}
is reduced since $P_{Df}$ is measured from the pulse peak to the $R$ baseline
adjacent to the peak. Accordingly, the true pulse height of the $D$ beam
itself should be larger than $P_{Df}$. The impact of this discrepancy on
$\Omega_{Di}$ can be estimated by inserting approximate values of
$F_{D}\approx0.4$, $F_{R}\approx0.02$, $\int I_{Di}\,da/\int I_{Df}%
\,da\approx0.92$ into Eq. (7)
\begin{align}
\Omega_{Di}  &  =\frac{P_{Di}}{P_{Df}}\nonumber\\
&  =\frac{\int F_{D}I_{Di}\,da}{\int F_{D}I_{Df}\,da-F_{R}\Delta}\nonumber\\
&  =0.924.
\end{align}
Consequently, the trial value $\Omega_{Di}$ is increased by a systematic error
of $\approx+0.004$ and an actual depletion of $\sim8\%$ would be
underestimated as $\sim7.6\%$.

The detector itself presents an additional potential source of error. A
systematic measurement error of the relative $P_{Di},P_{Df}$ pair values
potentially might arise from non-linearity of the detector output since these
values are acquired at different parts of the detector's photodiode response
curve because of the absence and presence of the fractional $R$ beam. However,
a comparable error would also be present in the $P_{Si},P_{Sf} $ pair values.
Accordingly, the observed $\left\langle P_{Si}/P_{Sf}\right\rangle
=1.00\pm0.01$, closely distributed about unity, provides verification of
detector linearity.

For individual trials, excursions of $P_{Si}/P_{Sf}$ significantly differing
from expected random variations of laser power output are indicative of
transitory mechanical instabilities of the apparatus.This has demonstrated by
temporarily re-positioning the $S$ beam spot to the center of the photodiode
and reducing the $S$ irradiance with the variable attenuator to $\sim40\%$.
The measured power fluctuations over 30 sec. intervals is a reasonably stable
$\pm0.5\%$ exceeded by the observed $\pm1\%$ fluctuations of $P_{Si}/P_{Sf}$
for which the $S$ beam spot is fractionally incident across the edge of the
photodiode. From this we conclude that the larger deviation of the
$P_{Si}/P_{Sf}$ trial ratios is primarily associated with increased detection
sensitivity to transient mechanical deflections of a fractionally intercepted
beam spot combined with a long optical beam path of $\sim4000mm$. An
equivalent mechanical stability sensitivity is expected for the accompanying
$P_{Di}/P_{Df}$. Accordingly, the modest dispersion of $\left\langle
P_{Si}/P_{Sf}\right\rangle =1.00\pm0.01$ is interpreted as a validation that
deviations of $P_{Di}/P_{Df}$ from unity are not significantly caused by
mechanical instability during the course of data acquisition.

As an additional means of eliminating systematic error, substitution of a
different detector (Hamamatsu S2386-8K 41 photodiode and a reverse bias
network) produced results equivalent to those obtained with the original
detector. This substitution is tantamount to verification of the experimental
results with a totally separate apparatus since quantum mechanics does not
allow for irradiance gain of the $D$ pulses upon coupling with the independent
$R$ beam regardless of how those $D$ pulses are generated.

The present apparatus is readily reproducible. In this regard, virtually all
components are widely available. An apparent exception is the diffractive beam
splitter, $DBS$ in Fig. 1, but any one of multiple alternative devices could
be used in substitution to split an incident beam into two beams of equal
irradiance. However, $G_{c}$ clearly emerges as a critical component since
local realism predicts that this grating, unlike an idealized grating, will
produce output beam occupation values in conflict with quantum mechanics.

In the present experiment, setting $P_{Di}=P_{Si}$, which provides for
intrinsic control values, suggests a useful application. This setting
effectively encodes the $D$ pulses with a higher wave intensity than those of
the $S$ pulses, but a receiver consisting of a simple energy-sensitive
detector measures the $D,S$ pulsed beam only as an unremarkable set of
constant irradiance pulses. However, a receiver additionally equipped with
restorative coupling decodes that higher wave intensity as an increased
irradiance on the equilibrated $D$ pulses. It is of some further practical
interest that the coupling of the $D$ pulses to the constant $R$ beam
constitutes a direct photonic amplification of those pulses.

\section{Conclusions}

From the perspective of local realism, exceptions to the invariance of
relative wave intensity (probability density) and energy flux density
(irradiance) are very subtle, and the formulation of the probabilistic
interpretation of quantum mechanics can be attributed to a presumed
universality of that invariance. That formulation necessarily abandons reality
and elevates the apparent wave-energy duality of photons to the status of the
central proposition of the probabilistic interpretation. The resultant compact
interpretation is compelling but nevertheless troubling given the abandonment
of reality and the imposition of non-locality. However, combined with Bell's
Theorem \cite{bell} and associated experimental results \cite{clauser}, that
interpretation is widely accepted.

The representation of local realism presented earlier [1, 2] provides a
plausible contradictory interpretation not constrained by Bell's Theorem
\cite{bell} and in agreement with experimental results \cite{clauser}. The
question of quantum mechanical completeness [10, 11] is addressed in that
representation by proposing that wave entities and energy quanta are
independent variables, a proposition experimentally tested here. The results
of this test show a statistically significant $\sim4\%$ energy gain consistent
with local realism and in conflict with quantum mechanics.

\end{document}